\documentclass[apj]{emulateapj}

\newcommand{\commentx}[1]{}

\renewcommand{\vec}[1]{\mbox{\boldmath$#1$}} 

\newcommand{\ra}[3]   
   {\makebox[1.5em][r]{#1}\makebox[1.5em][r]{#2} \makebox[2em][r]{#3}}

\newcommand{\hms}[3]  
   {${#1}^{\mathrm{h}}{#2}^{\mathrm{m}}{#3}^{\mathrm{s}}$}

\newcommand{\hmin}[2]  
   {\ensuremath{{#1}^{\mathrm{h}}{#2}^{\mathrm{m}}}}
   
\newcommand{\hours}[1]  
   {\ensuremath{{#1}^{\mathrm{h}}}}

\newcommand{\dms}[3]  
   {\ensuremath{{#1}\degree{#2}\arcminute{#3}\arcsecond}}

\newcommand{\dm}[2]  
   {\ensuremath{{#1}\degree{#2}\arcminute}}

\newcommand{\ukcmb}  
           {\ensuremath{\micro \kelvin_\mathrm{cmb}}}

\newcommand{\uk}  
           {\ensuremath{\micro \kelvin}}

\newcommand{\fdeg} 
           {\hbox{$.\!\!^{\circ}$}}

\usepackage{color}

\usepackage[citebordercolor={0 .5 .5}]{hyperref}
\usepackage{amsmath, amssymb, amsfonts}

\newcommand{\beq}{\begin{equation}}
\newcommand{\eeq}{\end{equation}}
\newcommand{\be}{\begin{equation}}
\newcommand{\ee}{\end{equation}}
\newcommand{\bea}{\begin{eqnarray}}
\newcommand{\eea}{\end{eqnarray}}
\newcommand{\bdi}{\begin{displaymath}}
\newcommand{\edi}{\end{displaymath}}

\def\lsim{\,\lower2truept\hbox{${<\atop\hbox{\raise4truept\hbox{$\sim$}}}$}\,}
\def\gsim{\,\lower2truept\hbox{${>\atop\hbox{\raise4truept\hbox{$\sim$}}}$}\,}

\shorttitle{Constraining bulk flow with kSZ}

\shortauthors{Mody \& Hajian}

\begin{document}

\title{One Thousand and One Clusters:\\
Measuring the Bulk Flow with the Planck ESZ and X-Ray Selected Galaxy Cluster Catalogs}

\author{
Krishnan~Mody,\altaffilmark{\dag,1}
Amir~Hajian\altaffilmark{\dag,2}
}
\altaffiltext{1}{Mathematics Department, Princeton University, Princeton, NJ, 08544, USA}
\altaffiltext{2}{Canadian Institute for Theoretical Astrophysics, University of
Toronto, Toronto, ON\ M5S~3H8, Canada}
\altaffiltext{\dag}{\url{ahajian@cita.utoronto.ca}}
\altaffiltext{\dag}{\url{kmody@princeton.edu}}

\begin{abstract}
We present our measurement of the ``bulk flow'' using the kinetic Sunyaev-Zel'dovich (kSZ) effect in the WMAP 7-year data.
As the tracer of peculiar velocities, we use  Planck Early Sunyaev-Zel'dovich Detected Cluster Catalog and a compilation of X-ray detected galaxy cluster catalogs based on ROSAT All-Sky Survey (RASS). We build a full-sky kSZ template and fit it to the WMAP data in W-band. Using a Wiener filter  we maximize the signal to noise ratio of the kSZ cluster signal in the data. 
We find no significant detection of the bulk flow, and our results are consistent with the $\Lambda$CDM prediction.
\end{abstract}


\section{Introduction}
Since the first claimed detection of large-scale streaming by \cite{Rubin/etal:1976} in Sc galaxies, the issue of coherent departures from uniform Hubble flow has been the source of much debate.  
The inflationary model predicts that the coherent, large scale  peculiar motion of matter caused by gravitational potentials, also called ``bulk  flow'',  is negligible in a $\Lambda$CDM universe  \citep{Strauss&Willick:1995}. This prediction has been tested in the last few decades and it has been the theme of much of the work in peculiar velocities. A known flow at small scales ( $< 60$ Mpc h$^{-1}$) is the motion of the Local Group towards a mass concentration of about $10^6 M_\odot$ known as the Great Attractor. This is closely associated and aligned with the observed CMB dipole \citep{Kogut/etal:1993, Jarosik/etal}. On large scales, \cite{Lauer&Postman} found a non-zero bulk flow at $80h^{-1}<R < 110h^{-1}$Mpc by using a full-sky peculiar velocity survey consisted of 119 Abel clusters. But a re-analysis of the data led to a reduced bulk flow in a different direction \cite{Hudson&Ebeling}. And at the same time \cite{Riess} used SNIa data and found no evidence of a bulk flow at similar scales.  \cite{Feldman/etal} used a compilation of peculiar velocity surveys and found a non-zero bulk flow at $R\sim 100h^{-1}$Mpc. 

All methods based on galaxy data are limited by our ability to observe and measure accurately at large distances. This limits the reach of these methods to scales of $\sim 100$Mpc. An independent measurement method on substantially larger scales can shed light on the matter and clarify the situation. The largest accessible probe to study the bulk flow to date is the kinetic Sunyaev-Zeldovich \citep{kSZ} effect due to the coherent motion of clusters of galaxies with respect to the rest frame of the Cosmic Microwave Background (CMB) radiation \citep{Haehnelt:1995dg}. Peculiar velocities of the electrons in the hot intracluster gas lead to a Doppler shift of scattered photons. The shift is proportional to the product of the line of sight component of the peculiar velocity and the electron density integrated along the line of sight through the cluster. A coherent flow causes an overall dipole seen through various optical depths of the clusters. This is a unique pattern that can be exploited in measuring the bulk flow.

Most recently \cite{Kashlinsky/etal:2008a} found a net dipole moment in the kSZ component measured in WMAP data that was consistent with a non-zero bulk motion at scales of $R \sim 575h^{-1}$Mpc. \cite{Keisler} repeated their analysis and found a negligible bulk flow in contradiction to the findings of \cite{Kashlinsky/etal:2008a}. \cite{Osborne} did an independent analysis of the kSZ in WMAP data and again found no significant velocity dipole. 

Although a lot of work has been done on this subject, the above summary clearly shows that there is no consensus on the existence of the bulk flow, its magnitude, depth or direction.  In this paper we use two different sets of cluster catalogs, based on SZ and X-ray detection of the clusters, to measure the bulk flow. 
Explaining the bulk flow within the framework of the $\Lambda$CDM model is difficult. 
Several theoretical models have been proposed to explain the possible existence of a large bulk flow at large scales in the framework of a DGP model \citep{Wyman&Khoury:2010} or by using other modified gravity models \citep{Afshordi/etal:2009, Khoury&Wyman}.

The rest of this paper is organized as follows: In Section \ref{sec:data}  we briefly describe the data sets used for
this analysis. In Section \ref{sec:method} we explain the details of the algorithm we use in constructing the full-sky kSZ templates, filtering, template fitting and error estimation. We present the results in Section \ref{sec:results}  and discuss its implications and systematics in Section \ref{sec:conclusion}. 
All through this paper we use the best-fit $\Lambda$CDM model of WMAP7 \citep{Larson/etal:2011} for the cosmological parameters.

\section{Data} \label{sec:data}
Our analysis uses two types of data sets: a data map and a cluster catalog. We use WMAP data with the highest resolution full-sky map to date. And as velocity tracers we use two independent catalogs of galaxy clusters; an SZ selected catalog and an X-ray selected catalog. Below we explain each of these data sets in more detail. 

\begin{figure}
\centering
\vspace{0.1cm}
\includegraphics[width = \columnwidth]{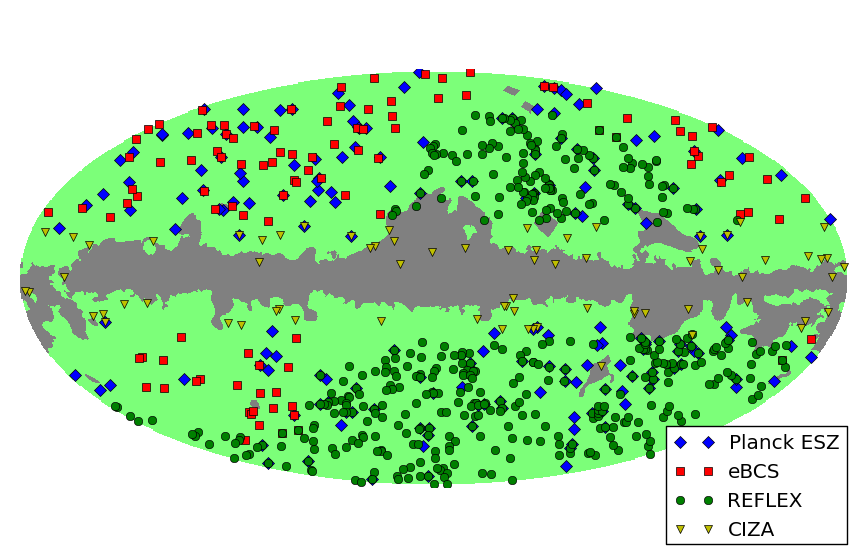}
\caption{The spatial distribution of the four catalogs and mask used in our analysis.}\label{fig:CatalogsInTheSky}
\end{figure}

\subsection{CMB Data}
We use co-added inverse-noise weighted data from seven single-year maps observed by WMAP at $94$ GHz (W-band).
The maps are foreground cleaned (using the foreground template model discussed in \cite{hinshaw/etal:2007}) and are at HEALPix \footnote{\url{http://healpix.jpl.nasa.gov}}  resolution 9 ($N_{\rm side}=512$). The WMAP data are signal dominated on large scales, $l < 548$ \citep{Larson/etal:2011} and the detector noise dominates at smaller scales. The noise in WMAP data is a non-uniform (anisotropic) white noise that varies from pixel to pixel in the map. Pixel noise in each map is determined by $N_{obs}$ with the expression
\be \label{eq:noise}
\sigma = \sigma_0/\sqrt{N_{obs}},
\ee
where $\sigma_0$ is the noise for each differencing assembly and can be found on the WMAP data products webpage on LAMBDA\footnote{\url{http://lambda.gsfc.nasa.gov/product/map/dr4/m_products.cfm}}. $N_{obs}$ is the number of observations at each map pixel which is directly proportional to the statistical weight, i.e. regions with larger number of observations have lower noise variances. $N_{obs}$ is included in the maps available from the LAMBDA website. 

In all of our analysis we use pixel masks to exclude foreground-contaminated regions of the sky from the analysis. We use a galactic mask which masks 19.30\% of the sky. We do not mask the point sources. 
 
\subsection{Cluster Catalogs: RASS X-ray Catalogs}

We use two sets of catalogs for the clusters of galaxies: three X-ray catalogs and one SZ catalog. There are 627 clusters in the X-ray catalogs and 189 clusters in the SZ catalog.
The \textit{Clusters in the Zone of Avoidance} (\textit{CIZA}) cluster catalog is an X-ray survey that identified clusters in the region where the magnitude of the galactic latitude is less than or equal to $20^{\circ}$. \textit{CIZA} used X-ray data from ROSAT All-Sky Survey (RASS) for its initial cluster candidate selection, and accepted or rejected candidate clusters according to optical and near infra-red (NIR) observations. This catalog has the locations of 73 clusters from the \textit{CIZA} survey, and the median redshift of these clusters is 0.07 (\cite{2002ApJ...580..774E}). 
The \textit{Extended ROAST Brightest Cluster Sample} (\textit{eBCS}) catalog is also an X-ray catalog. \textit{eBCS} was compiled from RASS data
and identified clusters in the northern hemisphere with galactic latitudes of magnitude greater than or equal to $20^{\circ}$; it is estimated to be $75\%$ complete. Total fluxes of clusters in this catalog are between $2.8\times10^{-12}$ and $4.4\times10^{-12}$ erg cm$^{-2}$ s$^{-1}$ (0.1 to 2.4keV). We have the locations of 107 clusters from the \textit{eBCS} catalog, and the median redshift of these clusters is 0.13. (\cite{2000MNRAS.318..333E}). 
The third X-ray cluster catalog we use is the \textit{ROSAT-ESO Flux Limited X-Ray} (\textit{REFLEX}) cluster catalog which covers 4.24 steradians in the southern sky. The sample is limited to
those clusters with an X-ray flux above $3 \times 10^{-12}$ erg s$^{-1}$ cm$^{-2}$ (0.1 to 2.4 keV), and it is estimated to be more than $90\%$ complete. We have the locations of 447 clusters identified by \textit{REFLEX}, and the median redshift of these clusters is 0.09 (\cite{2004A&A...425..367B}). 
\subsection{Cluster Catalogs: Planck ESZ}

\begin{figure*}
\centering
\vspace{0.1cm}
\includegraphics[width = \textwidth]{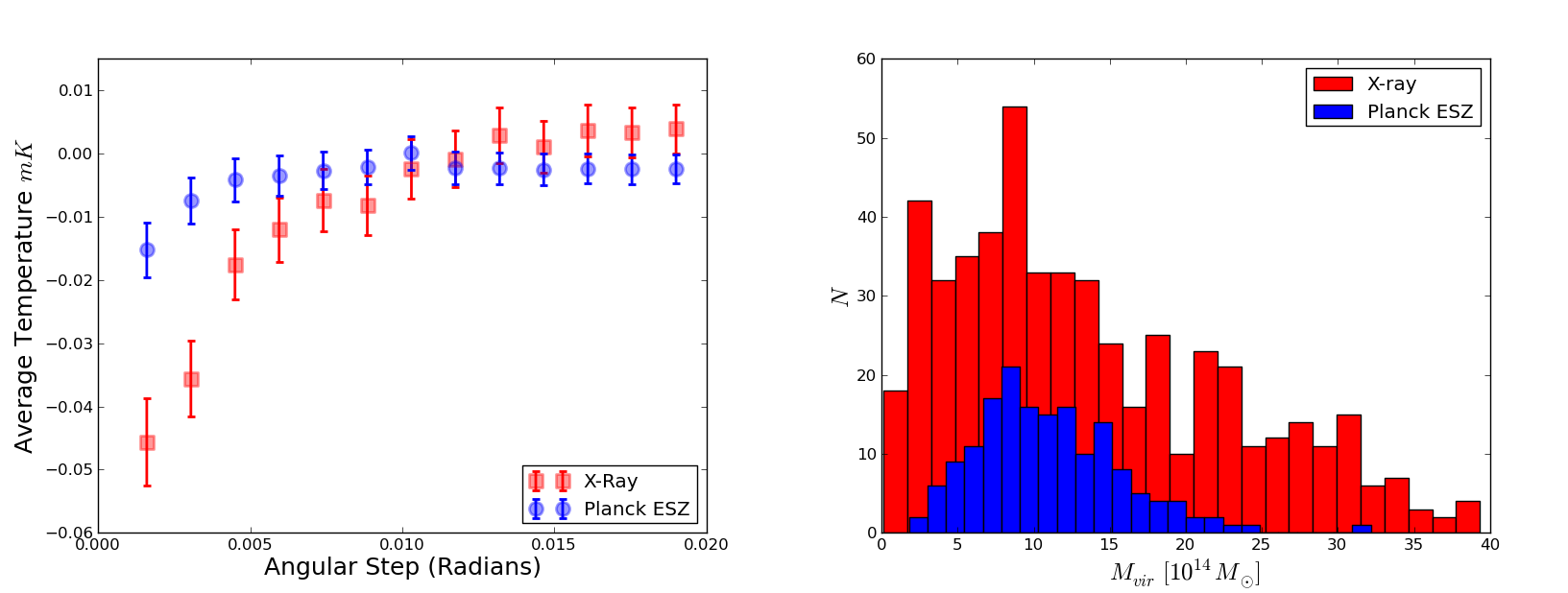}
\caption{\textit{Left}) The stacked cluster profiles in \textit{Planck ESZ} and the X-Ray catalogs in rings of width 0.00145 Radians around each cluster. The error bars are given by the standard deviation of this quantity for all the clusters in the catalog.  Only clusters with absolute latitude $|b| > 25^\circ$ are used in the stacking. \textit{Right}) The distribution of $M_{vir}$ in \textit{Planck ESZ} and the X-Ray catalogs.}\label{fig:together}
\end{figure*}

\textit{Planck} is mapping the whole sky at a few arcminute resolution and will eventually produce an SZ detected cluster catalog of a few thousand galaxies. What is currently available, hereafter \textit{Planck ESZ} catalog, is an isotropic all-sky catalog and has 189 clusters with a median redshift of 0.546. 
Figure \ref{fig:CatalogsInTheSky} shows the spatial distribution of the clusters of galaxies in the four catalogs discussed above, and Figure \ref{fig:together} compares the stacked cluster profiles. 

\section{Generating and fitting full-sky kSZ templates} \label{sec:method}

We generate a full-sky kSZ template due to the bulk flow and measure the components of the bulk flow velocities by fitting the template to the data in real space. The kSZ template is built using an indicator of the mass of the cluster. We don't have a direct mass estimate in the catalogs, so we use scaling relations to derive the masses based on other properties of the clusters, in particular their luminosities and their integrated Compton-$Y$ factors depending on the catalog. 

\subsection{Cluster Parameters: X-Ray Catalogs}
In each of the X-ray catalogs, we have the luminosity of each cluster, and we use the luminosity to find the virial mass and radius for each cluster. To do so, we
first find $M_{500}= \frac{4\pi}{3}[500 \rho_c(z)]R_{500}^3$, the mass of the cluster contained within $R_{500}$, the radius within which the mean overdensity of the cluster is 500 times the critical density of the universe, $\rho_c(z) = 2.775 \times 10^{11} E^2(z) M_{\odot}$ Mpc$^{-3}$. We use the scaling relation of \cite{2009ApJ...692.1033V} to compute $M_{500}$ from the X-ray luminosities
\bea \label{eq:L-M:fit}
\ln L_X &=& (47.392 \pm 0.085) + (1.61\pm0.14)\ln M_{500}\nonumber \\ 
    &+&(1.850\pm0.42)\ln E(z)-0.39\,\ln(h/0.72)\nonumber \\ 
    & &\pm (0.396\pm 0.039),
\eea
where $L_X$ is the X-ray luminosity in units erg s$^{-1}$, $M_{500}$ is in units of $M_{\odot}$ and $E(z) = [\Omega_m(1+z)^3 + \Omega_\Lambda]^{\frac{1}{2}}$ is the expansion history of the universe (\cite{2009ApJ...692.1033V}). 
From $M_{500}$ we simultaneously calculate the virial mass, $M_{vir}$, i.e., mass enclosed within the virial radius, $R_{vir}$,  and the concentration
parameter,$c$ , using the system of equations
  \bea   \label{eq:system}
    M_{vir} &=& \frac{4\pi}{3}[\Delta_c(z)\rho_c(z)]R_{vir}^3, \\  \nonumber
    M_{500} &=& M_{vir}\frac{m(cR_{500}/R_{vir})}{m(c)},  \\ \nonumber
    c &=& \frac{5.72}{(1+z)^{0.71}}\left( \frac{M_{vir}}{10^{14}h^{-1}M_{\odot}}  \right)^{-0.081},
  \eea
where $m(x) = \ln(1+x) - \frac{x}{1+x}$, $c$ is taken from  \cite{Duffy/etal:2008} based on the N-body simulations with the
WMAP five-year cosmological parameters and $\Delta_c(z)$ is a function of $\Omega_m$ and $\Omega_\Lambda$ \citep{1998ApJ...495...80B}
\be
\Delta_c(z) = 18\pi^2 + 82[\Omega(z) - 1] -39[\Omega(z) - 1]^2,
\ee
and $\Omega(z) = \Omega_m(1+z)^3/E^2(z)$.   For reference, $2 R_{500}$ approximates $R_{vir}$ (\cite{2011ApJS..192...18K}).
We use the concentration parameter to define the scaling radius, namely
  \beq
    r_s = \frac{R_{vir}}{c}.
  \eeq
We numerically solve the system of equations (\ref{eq:system}) using  FuncDesigner library for Python\footnote{http://openopt.org/FuncDesignerDoc}. 

\subsection{Cluster Parameters: SZ Catalog}

\textit{Planck ESZ} contains the integrated compton-Y at the cluster position and within $5 R_{500}$. Here the compton-Y
parameter is given by
  \beq
    \frac{\sigma_T}{m_e c^2} \int P\cdot dl,
  \eeq
where $\sigma_T$ is the Thomson cross section, $m_e$ is the rest mass of an electron, $P$ is the intracluster medium thermal electron pressure, and the integral is taken along the line of sight and over the area of the cluster in question. To find $M_{500}$, we use the scaling relation
    $Y_{R_{500}} = 0.55 Y_{5R_{500}}$
which gives the integrated compton-Y at X-ray position and within $R_{500}$. Furthermore, we use the scaling relation 
  \beq
    \frac{Y_{R_{500}}}{E(z)^{\frac{2}{3}}} = 10^A \left( \frac{M_{500}}{10^{14}M_{\odot}}  \right)^B,
  \eeq
 where $A=-4.213$, $B = 1.72$ (\cite{2011arXiv1101.2026P}). The distribution of $M_{vir}$ in the SZ and X-Ray catalogs is shown in Figure \ref{fig:together}.
Then we use the system of equations (\ref{eq:system})  to find $M_{vir}$ and $R_{vir}$.

\subsection{Cluster Profiles}

We use a $\beta$-model for the cluster profiles.  In this model, the number density of electrons in a cluster is given by

\beq
    n_e(r) = n_{e0}\left(  1 + \frac{r_s^2}{r^2}  \right)^{-3\beta/2},
  \eeq 
  with
  \beq
    n_{e0} = \frac{N_e}{4 \pi r_s^2 \left(c - \arctan c \right)},
  \eeq  
and we take $\beta = 2/3$ to describe the X-ray surface brightness profile of observed clusters \citep{mann&mann}. $r$ is defined within the virial radius of the cluster. We calculate $N_e$, the total number of electrons in each cluster, according to 
   \begin{equation}
      N_e = \left(\frac{1 + f_H}{2m_p} \right)f_{gas}M_{vir}.
  \end{equation}
Here $f_H = 0.76$ is the hydrogen fraction, $\displaystyle f_{gas} = \Omega_b/\Omega_m = 0.0168$,
and $m_p$ is the proton mass. 
Taking the line of sight integrals, the optical depth of a cluster in the $\beta$-model is given by
  \beq
    \tau(\theta) = 2\sigma_T \frac{r_s n_{e0}}{\sqrt{1 + \frac{\theta^2}{\theta_c^2}}} \tan^{-1} \sqrt{\frac{c^2 - \frac{\theta^2}{\theta_c^2}}{1 + \frac{\theta^2}{\theta_c^2}}}.
  \eeq 
In order to check the effect of the assumed cluster profile on our results, we do our analysis using the \cite{NFW} (NFW) model  as well. The electron density in this model is given by
 \beq
    n_e(r) = \frac{n_{e0}}{\frac{r}{r_s} \left( 1 + \frac{r}{r_s} \right)^2},
  \eeq
  with 
  \beq
    n_{e0} = \frac{N_e}{4\pi r_s^3 \left( \ln(1 + c) - \frac{c}{1 + c} \right)}.
  \eeq
And the optical depth is given by
\beq
    \tau(\theta) = 
      \begin{cases} 
        \frac{2 r_s n_{e0}}{x^2 - 1}\left( 1 - \frac{2}{1 - x^2} \tanh^{-1} \sqrt{\frac{1 - x}{1 + x}} \right) & x < 1\\ \nonumber
        \frac{2 r_s n_{e0}}{3} & x = 1\\ \nonumber
        \frac{2 r_s n_{e0}}{x^2 - 1}\left( 1 - \frac{2}{1 - x^2} \tan^{-1} \sqrt{\frac{x - 1}{1 + x}} \right) & x > 1\\
      \end{cases}
  \eeq
where $x \equiv \frac{r}{r_s}$ (\cite{2000ApJ...534...34W}).
The key difference between the two models is that the NFW profile places most electrons at the central region of the cluster whereas the $\beta$-model is more extended throughout the cluster and beyond. In both cases we cut off the model at $r_{vir}$. In practice we project the cluster profiles onto a pixelized sphere. To pixelize our models, we calculate
  \begin{equation}
    \tau_{total} = D_{A}^{-2}\sigma_T N_e,
  \end{equation}
with $\sigma_T$ the Thomson scattering coefficient. Every cluster is built within its $\theta_{vir}$. For each pixel in the cluster,  $\tau(\theta)$ is computed and assigned to that pixel. Pixel values are normalized in each cluster by
  \begin{equation}
    \frac{\tau_{total}}{\sum_{i} \tau_i\Omega_{pix}}.
  \end{equation}
Here the sum is taken over all the pixels within the given cluster,
and $\Omega_{pix}$ is a constant for the entire map given by $\displaystyle\frac{4\pi}{n_{pix}}$. Our normalization is such that
  \begin{equation}
    \sum_{i} \tilde{\tau}_i\Omega_{pix} = \tau_{total},
  \end{equation}
where $\tilde{\tau}_i$ is the normalized value of each pixel. The typical opitcal depths that we obtain using this algorithm are of the order of $10^{-3}$. 

\subsection{Template fitting}
\noindent The optical depth templates explained above are used to create KSZ templates for unit velocities in the $x$, $y$ and $z$ directions, $[1,0,0]$ km/s, $[0,1,0]$ km/s, $[0,0,1]$ km/s respectively by multiplying each pixel by
\beq
  -T_{cmb} \frac{\hat{n} \cdot \vec{v}}{c},
\eeq
where $\hat{n}$ is the vector pointing to the pixel in question, and $T_{cmb}$ is the temperature of the CMB. 
We also use a template for the thermal component of the SZ (tSZ) signal \cite{tSZ} in our analysis
\bea \label{eqn:tSZ}
  \Delta T^{tSZ} &= \left( \frac{x(e^x + 1)}{e^x - 1} - 4  \right) T^{CMB}y, \\ \nonumber
  y &= \frac{k_B}{m_e c^2}T_e \tau, \\ \nonumber
  x &= \frac{h\nu}{k_BT},
\eea
where $k_B$ is Boltzman's constant, $m_e$ is the mass of an electron, $h$ is Planck's constant and $\nu$ is the frequency of the CMB map (94GHz). We assume a uniform temperature for the electron gas in the clusters to be $T_e = 4 \cdot 10^7 K$. 
Before we measure for a bulk velocity, we apply a Wiener filter to our templates and data. To do so, we take as the denominator of our filter the CMB power spectrum plus white noise and we take as numerator of our filter the power spectrum of a kSZ template with velocity $[0, 0, 1200]$ $km/s$  \citep{Kashlinsky/etal:2008a}. Our Wiener filter is based on the theory power spectrum and is not derived from the data unlike the Wiener filter of \cite{Kashlinsky/etal:2008}. For a study of diffenet filters in this context see \cite{Osborne}.

Then we use the least square minimization to fit the above templates to the data
\be
  \chi^2 = \left(\mathbf{D} - \mathbf{\alpha}\cdot\mathbf{T}\right)^t \mathbf{C}^{-1} \left(\mathbf{D} - \mathbf{\alpha}\cdot\mathbf{T}\right),
\ee
where $\mathbf{D}$ is the CMB map and $\mathbf{\alpha}\cdot\mathbf{T} = v_x T_x^{ksz}+v_y T_y^{ksz} + v_z  T_z^{ksz} + A T^{tSZ}$ is the tSZ plus kSZ fit to the map. We take the covariance matrix to be diagonal and determined by the pixel noise in WMAP. This is a good approximation for the present data because the cluster sample is sparse and we are dominated by the pixel noise on small angular scales of interest. We use Monte-Carlo simulations to test our method.
The result of the fit is the four-vector $\mathbf{\alpha}$ that contains the $x$, $y$ and $z$ components of the bulk flow and the tSZ template amplitude. Results are presented in Section \ref{sec:results}.

In order to assess the statistical significance of the results, we use Monte-Carlo simulations of the CMB sky. Simulated maps have three components
\be\label{eq:sim}
\Delta T(\hat{n}) = \left(\Delta T_{CMB}(\hat{n})+\Delta T_{tSZ}(\hat{n})\right)\otimes B(\hat{n}) + N(\hat{n}),
\ee
where $\Delta T_{CMB}$ is a realization of the Gaussian CMB field, $\Delta T^{tSZ}$ is the thermal component of the SZ signal simulated as described in eqn. (\ref{eqn:tSZ}), $N(\hat{n})$ is the pixel noise and $\otimes B(\hat{n})$ means convolved with the proper beam of the experiment. 

We make 100 realizations of the CMB sky using {\tt synfast} routine of HEALPix\footnote{We use {\tt healpy}, the Python version of HEALPix.} with the underlying theory power spectrum computed with CAMB\footnote{\url{http://camb.info}} using  the concordance model. 
The maps are then convolved with WMAP beam for W band.  Noise realizations are added to the beam convolved maps in the end. Noise maps are simulated using eqn.(\ref{eq:noise}) with $\sigma_0 = 6.549$ mK  for the W-band noise.   We test our simulations by comparing their average power spectra with the data power spectrum.
\begin{figure}
\centering
\vspace{0.1cm}
\includegraphics[width = \columnwidth]{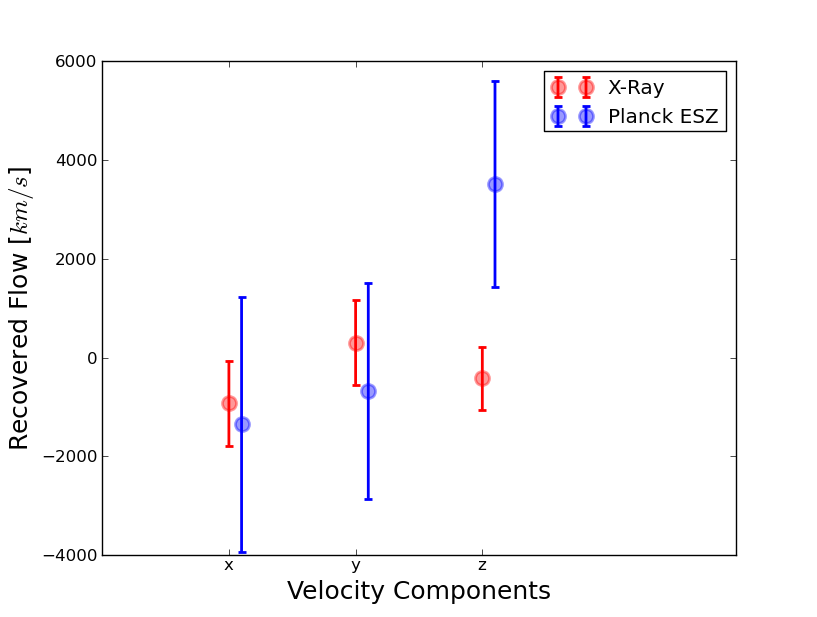}
\caption{Bulk velocity components measured using the two catalogs. Our results do not show a significant bulk flow.   The larger errorbars on the Planck catalog measurements are due to the smaller number of objects in the catalog.     }\label{fig:results}
\end{figure}
\section{Results}\label{sec:results}
For each cluster catalog we perform the fits using two sets of templates: templates based on the NFW model for the cluster profiles and templates using the $\beta$-model. Using the NFW model, we obtain the bulk velocity [-924, 305, -421] $\pm$ [864, 852, 644] $km/s$ in the combined catalog of X-ray clusters and [-1350, -668, 3520] $\pm$ [2581, 2187, 2085] $km/s$ for the SZ catalog. The results for the NFW profile are shown in Fig. \ref{fig:results}.
 Using a $\beta$-model, does not significantly change the results. For the $\beta$-model described above we obtain the bulk velocity of [-1110, 123, -391] $\pm$ [872, 901, 689] $km/s$ for the X-ray catalog and [-2940, -1120, 4840] $\pm$ [3512, 3443, 2441] $km/s$ for the SZ catalog. In order to measure the bulk flow at different redshifts, we divided the X-ray catalog into three catalogs at redshifts $z < .068$, $0.068 <z < 0.13$ and $z>0.13$. The redshift bins are chosen so that all bins have roughly the same number of clusters in them. We repeated our analysis for each of these samples. The bulk flow was consistent with zero in all three redshift bins.


\section{Summary and Conclusion}\label{sec:conclusion}
We use cluster catalogs and the highest resolution maps of the WMAP data (W-band) to measure the bulk flow using the kSZ effect. We do not detect a significant bulk flow. The best constraint we get on the bulk flow velocities is from the RASS X-ray catalog, assuming an NFW profile for the clusters. The result is shown in Fig. \ref{fig:results}. Using a $\beta$ model leads to similar results. This is due to the size of the WMAP beam which makes the details in the cluster profiles indistinguishable at this resolution. The results based on the Planck ESZ catalog have large uncertainties due to the smaller number of the clusters in the catalog. 
Our results agree with the results of \cite{Osborne, Keisler} and contradict the findings of \cite{Kashlinsky/etal:2011} where they report a significant detection of a bulk flow inconsistent with zero. It is important to have as many independent statistical methods as possible to measure and constraint the bulk flow velocities. Future cluster catalogs and CMB maps will help us tighten these constraints.

This analysis can be improved by using larger cluster catalogs such as the future X-ray selected catalog of eRosita All-Sky Survey \citep{Cappelluti}, and the next releases of Planck SZ selected clusters combined with the SZ detected clusters of ACTPol \citep{ACTPol} and SPTpol \citep{sptpol}. However, using more clusters alone will not improve the analysis much; one needs to use a higher resolution CMB map as well to increase the signal to noise ratio in the measurement. The only full-sky maps of the CMB at higher resolution in the near future are going to be Planck maps. Using Planck data with future cluster catalogs will help us put tighter constraints on the bulk flow \citep{Mak}.  As the data get better, it is important to improve the theoretical models used for making the templates. For example second order effects like the  changes in the SZ cluster brightness, flux and number counts induced by the motion of the Solar System  \citep{Chluba} and the relativistic corrections to the kSZ signal \citep{relativistickSZ} need to be taken into account. 

\begin{acknowledgments}
We would like to thank Nick Battaglia, Niayesh Afshordi and Neelima Sehgal for very useful discussions and feedback throughout this work, and Viviana Acquaviva for her comments on the manuscript. We thank Mike Nolta for providing us with the galactic mask and the scaling relation for the $R_{500}$ in the Planck model. Finally, we thank David Spergel whose guidance was integral to the progress of the project. We acknowledge the use of the Legacy Archive for Microwave Background Data Analysis (LAMBDA). Support for LAMBDA is provided by the NASA Office of Space Science. Some of the results in this paper have been derived using the HEALPix \citep{gorski/etal:2005} package.

\end{acknowledgments}

\end{document}